\documentclass[traditabstract]{aa}
\usepackage[utf8]{inputenc}
\usepackage{microtype}
\usepackage{paralist}
\usepackage{amsmath}
\usepackage{natbib}
\usepackage{graphicx}
\usepackage{epsfig}
\usepackage{txfonts}
\usepackage{color}
\usepackage{multirow}
\usepackage{amssymb}
\usepackage{epsfig}
\usepackage{xspace}
\bibpunct{(}{)}{;}{a}{}{,} 
\usepackage[plainpages=false]{hyperref}

\newcommand{\vela}{Vela~X$-$1\xspace}
\newcommand{\rt}{Rayleigh–Taylor\xspace}
\newcommand{\kh}{KHI\xspace}
\newcommand{\ig}{\textsl{INTEGRAL}\xspace}
\title{Witnessing the magnetospheric boundary at work in \vela.}
\author{V.\,Doroshenko\inst{1}, A.\,Santangelo\inst{1}, V.\,Suleimanov\inst{1,2}}	
\institute{Institut für Astronomie und Astrophysik, Sand 1, 72076 Tübingen, Germany\and
Kazan State University, Kremlevskaya 18, 420008, Kazan, Russia}

\begin{document}

\bibliographystyle{aa}

\abstract{We present an analysis of the \vela ``off-states'' based on
\emph{Suzaku} observations taken in June 2008. Defined as states in which the
flux suddenly decreases below the instrumental sensitivity, these
``off-states'' have been interpreted by several authors as the onset of the
``propeller regime''. For the first time ever, however, we find that the
source does not turn off and, although the flux drops by a factor of 20 during
the three recorded ``off-states'', pulsations are still observed. The spectrum
and the pulse profiles of the ``off-states'' are also presented. We discuss
our findings in the framework of the ``gated accretion'' scenario and conclude
that most likely the residual flux is due to the accretion of matter leaking
through the magnetosphere by means of Kelvin-Helmholz instabilities (KHI).}

\authorrunning{V. Doroshenko et al.}
\maketitle

\section{Introduction} \vela is a persistently active high-mass X-ray binary
system (HMXB) consisting of a massive neutron star (1.88$M_\odot$,
\citealt{velamass}) and a B0.5Ib type super giant HD~77581 with mass of $\sim23M_\odot$ and
radius of $\sim30R_\odot$ \citep{kerkwijk}.
The average X-ray luminosity of the pulsar, $L_{\rm
X}\sim4\times10^{36}{\rm\,ergs\,s}^{-1}$, is explained well by the mass-loss
rate of the optical companion of $\sim10^{-6}M_\odot\,\mathrm{yr}^{-1}$
\citep{nagase86}, assuming a simple wind model \citep{castor} and the
observed terminal wind velocity of $\sim1100$\,km\,s$^{-1}$
\citep{watanabe06}. The neutron star is eclipsed by the optical companion
every orbital cycle of $\sim8.964$\,d \citep{kerkwijk}. The spin period of the
neutron star, $P_s\sim283.5\,\text{s}$ \citep{rappaport_vela}, has remained
almost constant since this discovery. The X-ray spectrum of \vela is
described well by a cutoff power law \citep{nagase86}. Cyclotron resonance
scattering features have been reported at $\sim25$ \citep{Makishima:1992p3220}
and $\sim50-55$\,keV \citep{kendziorra_vela}, although the feature at 25\,keV
is still debated \citep{orlandini06}.

The source is strongly variable with an average X-ray luminosity of
$\sim4\times10^{36}\mathrm{erg\,s}^{-1}$ (assuming a distance of
$\sim2$\,kpc, \citealt{nagase89}). Aside from the usual flaring activity
similar to the one observed in other wind-accreting pulsars, abrupt
``off-states'', in which the source becomes undetectable for several pulse
periods, and ``giant flares" characterized by an increase in the flux up to a
factor of 20 have been observed \citep{kreyken_vela}. Both phenomena are not
unusual for \vela, although probably neither has a periodic nature
\citep{kreyken_vela}. The giant flares of \vela have been compared to those
seen in Super Fast X-ray Transients (SFXTs, \citealt{Walter:2007p419}), and
may be connected to the clumpy structure of the wind. In this scenario the
wind velocity and the density fluctuations are responsible for the luminosity
swings: flares would be associated with the accretion of particularly massive
wind clumps, whereas ``off-states'' can be explained with the onset of the
centrifugal barrier \citep{Illarionov:1975p2044}, triggered by a drop in wind
density/velocity \citep{kreyken_vela}.

An alternative ``gated accretion'' scenario to explain the flaring activity of
SFXTs was proposed by \citet{grebenev07} and \citet{Bozzo:2008p2039}, who suggested that the observed
luminosity swings may be associated to the transition between different
accretion regimes, i.e. to the different ways the plasma enters the
magnetosphere. It is the interaction of the rotating neutron star's
magnetosphere with the plasma that mediates the accretion rate. When the
magnetospheric boundary is strong, the matter accumulates close to the
magnetospheric boundary. If the growth of the magnetospheric instabilities
renders the barrier transparent, the accumulated matter can accrete, thereby producing
a bright flare. Observational evidence for this scenario was reported by
\cite{Bozzo:2008p2039} and \cite{grebenev10}.

In this work we report on the analysis of the June~2008 \emph{Suzaku}
observation of \vela. We focus on the three ``off-states'' detected in the
observation. To our knowledge it is the first time that this type of activity
has been observed with an instrument sensitive enough to constrain the flux in
the ``off-state'', to detect pulsations, and to perform the spectral and pulse
profile analysis.

We conclude that the accretion still proceeds during the ``off-states'',
although at a much lower rate. We discuss our result in the framework of the ``gated
accretion'' scenario \citep{burnard83, Bozzo:2008p2039} mentioned above. The
abrupt decrease in the accretion rate during the ``off-states'' may then be
explained if the magnetospheric boundary becomes stable with respect to \rt
instability, but some matter still leaks because the magnetosphere is still
unstable with respect to Kelvin-Helmholz instabilities (KHI).

\section{Data analysis and results} 
\begin{figure*}[t]
	\centering
	\includegraphics[width=\textwidth]{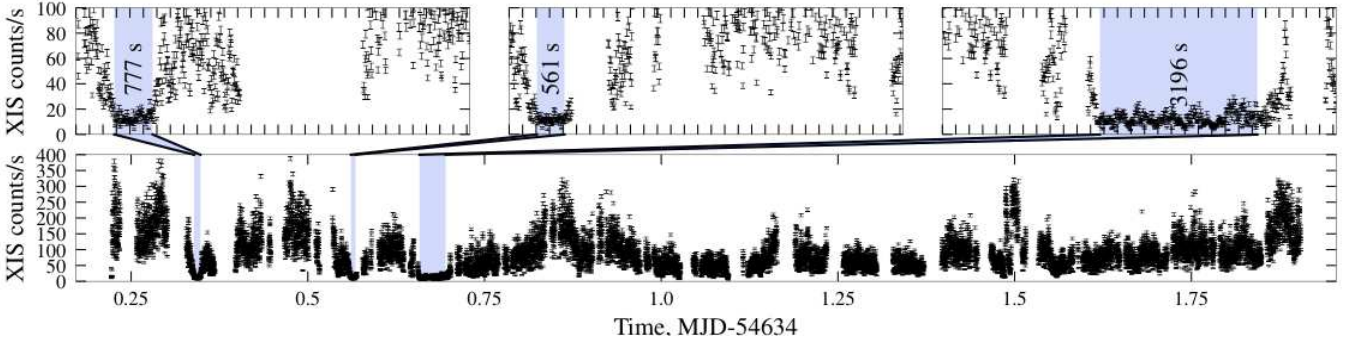}
	\caption{Observation-long lightcurve in the 0.4-12\,keV energy range using data from all XIS units is shown in the bottom panel. The upper panels show close-up views of the three detected off-states. Here, the time axis is ticked every pulse period.}
	\label{fig:obslc}
\end{figure*}

The observation we rely on is a
$\sim100$\,ks long \emph{Suzaku} observation (ID 403045010), performed on
June~17-18, 2008, about 1.6\,d after the eclipse and close to the periastron
passage of the source (orbital phase $\sim0-0.16$). The data was
reduced using the HEADAS~6.9 with CALDB version 20100812. \paragraph{Timing
analysis.} To improve the quality of the statistics, data from all three XIS units were
combined. The lightcurve of the observation in the range 0.4-12\,keV is
presented in Fig.~\ref{fig:obslc}. Three ``off-state'' episodes, shown in the
upper panels of Fig.~\ref{fig:obslc}, are observed. As reported in literature
\citep{inuoe,Lapshov92,kreyken_vela_99,kreyken_vela}, during the
``off-states'' the sources' flux drops abruptly and recovers after several
pulse periods. Marginal evidence of a residual pulsed emission was reported
by \cite{inuoe} based on \emph{Tenma} data. With the unprecedented sensitivity
of \emph{Suzaku} not only we do unambiguously confirm these findings but we can
also study the ``off-states'' in detail.

\begin{figure*}[t]
	\centering
		\includegraphics[width=\textwidth]{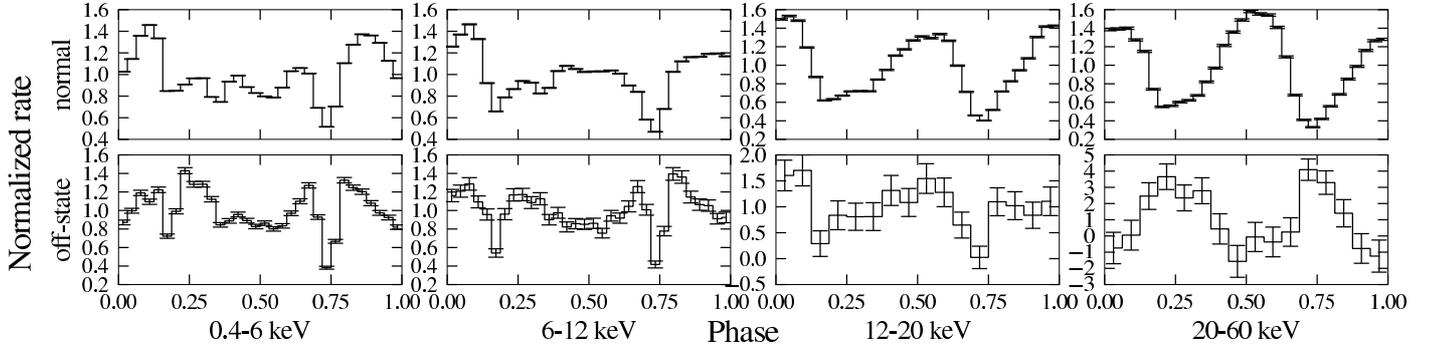}
	\caption{Pulse profiles in four different energy ranges are shown for the normal (upper panels) and the ``off-state'' (bottom panels).}
	\label{fig:pp}
\end{figure*}

Using the phase-connection technique \citep{staubert2009} and assuming the
ephemeris by \cite{kreyken_vela}, we determined the pulse period to be
$P_s=283.473(4)$. All uncertainties quoted are at $1\sigma$ confidence level
unless stated otherwise. The marginal evidence of spin-up is not statistically
significant and may be attributed to the uncertainty of the orbital
parameters. No change in the pulse period in ``off-states'' could be measured.
Based on the obtained timing solution, we constructed energy-resolved pulse
profiles for the entire observation and for the ``off-states'' by combining all
three episodes. As can be seen in Fig.~\ref{fig:pp}, the pulse profiles
vary significantly with both energy and luminosity. Very remarkable is
the change around phase $\sim 0.75$ at hard energies (20-60~keV) between the
normal and the ``off-state'' profile.

\begin{figure}[t]
	\centering
		\includegraphics[scale=0.95]{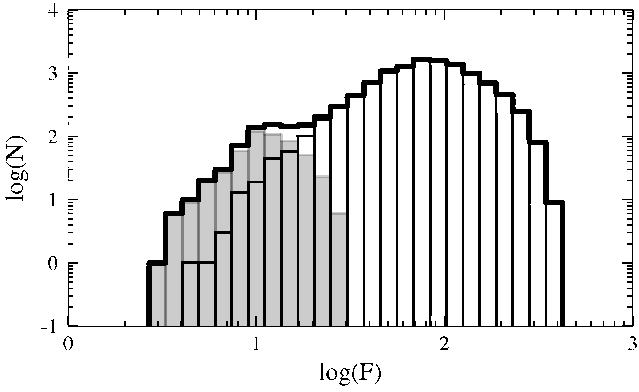}
	\caption{Histogram of the measured count-rate of the XIS lightcurve (solid line). The same is shown for the off-states (shaded) and for the rest of the lightcurve separately.}
	\label{fig:hist}
\end{figure}

We also investigated the flux distribution of the source following the
approach of \cite{fuerst}, who shows that the flux distribution is
approximately lognormal based on \ig data. Although the lognormal distribution
generally describes our data, an excess appears at low count rates
(Fig.~\ref{fig:hist}). This excess is due to the ``off-states''. The flux
distribution of the ``off-states'' is still approximately lognormal
(Fig.~\ref{fig:hist}) but differs considerably from the distribution of the
rest of the lightcurve. It is the ``off-states'' component that mostly
contributes to the low-countrate flank seen in the overall flux histogram.
Implications of this finding are discussed below. \paragraph{Spectral
analysis} We first analyzed the average spectrum of the entire observation to
establish a baseline for the analysis of the ``off-state'' data. Several
phenomenological continua, based on the models reported in literature for
\vela, were used to fit the average spectrum of the source. None of these
models was able to describe the broad-band 0.4-70\,keV spectrum. In
particular, the spectrum below 5\,keV is poorly described by cut-off power-law
models.

\begin{figure}[t]
	\centering
		\includegraphics[scale=0.95]{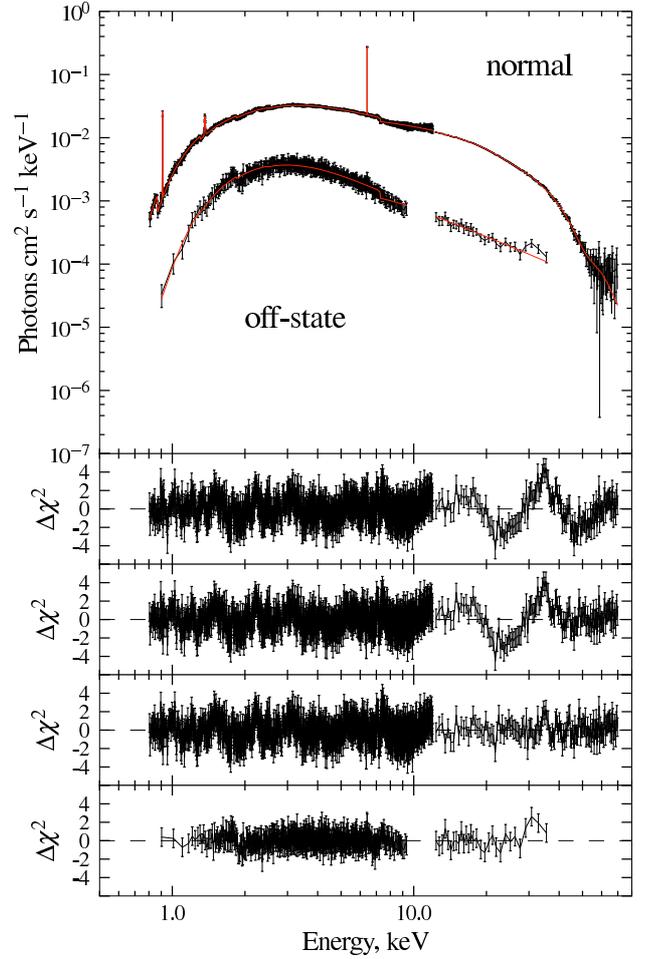}
	\caption{Unfolded spectrum of the normal and ``off states".
	For the normal spectrum XIS0 and HXD PIN data were used. For the ``off'' spectrum data from all XIS units were combined to obtain a larger statistics. Best-fit residuals are also shown from top to bottom for: the normal state without CRSF; with the inclusion of a CRSF at 55\,keV; with the inclusion of two CRSFs, and, eventually, for the ``off-state'' spectrum.}
	\label{fig:spe}
\end{figure}

We used two components to model the continuum, combining a Comptonization model by
\cite{Titarchuk:1996p2241} and a power law. Photoelectric absorption at lower energies,
a number of emission lines, and an iron absorption edge at $\sim7.26$ \citep{nagase86} were
also necessary to fit our data. Two CRSF harmonics were also required by the fit and
were modeled using a multiplicative Gaussian profile. The best-fit
parameters are summarized in Table~\ref{tab:alspe}. The
quality of the statistics is significantly lower for the ``off-state'' data, so a simpler model was
used in this case: an absorbed Comptonization model with the addition of an iron
absorption edge. No other spectral features were required by the fit. The results of
the best fit are presented in Table~\ref{tab:alspe}, and the best-fit spectra and the fit residuals are shown in
Fig.~\ref{fig:spe}. The average absorption-corrected flux in the 0.4-70\,keV energy
range was $\sim3.8\times10^{-9}{\,\rm erg\,s}^{-1}$ for the complete observation and
$\sim5\times{10}^{-10}{\,\rm erg\,s}^{-1}$ for the ``off-state'' spectrum. Results of a
more detailed spectral analysis, including phase-resolved spectra, will be published
elsewhere.
\begin{table*}
	\begin{center}
	\begin{tabular}{llllllllllllll}
\hline
\hline
        & $T_{0,c}$ &  $kT_c$ &                      $\tau_c$ & $A_{\rm comp}$ & $N_{\rm H}$ & $\Gamma$ & $A_\Gamma$ & $E_{\rm cyc,25}$ & $\sigma_{\rm cyc,25}$ & $\tau_{\rm cyc,25}$ & $E_{\rm cyc,50}$ & $\sigma_{\rm cyc,50}$ & $\tau_{\rm cyc,50}$\\
\hline
 ``On'' &   0.98(2) & 7.97(3) &                         15(2) &        0.07(2) &     1.45(8) &   3.2(3) &    0.07(2) &          26.6(9) &                  7(1) &              0.3(1) &            55(3) &                 13(4) &              1.7(1)\\
``Off'' &   0.83(8) & 21.0(1) & $\displaystyle7^{1.5}_{-4.4}$ &     $\le 0.03$ &      1.4(1) &          &            &                  &                       &                     &                  &                       &                    \\
\hline
	\end{tabular}
	\end{center}
	\caption{Best-fit parameters of the normal and ``off-states" spectra with uncertainties at $1\sigma$ confidence.}
	\label{tab:alspe}
\end{table*}

It is interesting to note that the ``off-state'' spectrum differs
considerably from the spectrum observed during the eclipses when it is dominated by
emission-lines \citep{watanabe06} originating in the surrounding plasma illuminated by the X-rays emitted by the eclipsed pulsar.
\section{Interpretation and discussion}
First, we would like to summarize the observed properties of the source in its ``off-state''
\begin{compactitem}
	\item The flux drops by a factor of ten or more on a timescale 
comparable to the pulse period. The source remains ``off'' for several pulse
periods, and then the flux is restored to the previous level on the same
short timescale.
	\item \vela is observed to pulsate in the ``off-state''. No pulse frequency change has been detected in the ``off-state''.
	\item A drastic change in the shape of the pulse profile shape appears at high energies. The narrow dip at pulse phase $\sim0.75$, seen in the profiles of the normal state at all energy ranges as well as in the ``off-state'' profiles at lower energies, is substituted by a prominent peak in the ``off-state'' profile at hard energies (20-60)~keV. This leads to a significant increase in the fraction of the pulsed emission in the hard energy range for the ``off-states''.
	\item The flux distributions of the normal and ``off-states'' are significantly different. The overall distribution is composed of two
	approximately lognormal peaks.
	\item Although poorly constrained at high energies, the spectrum changes
significantly during the ``off-state''. The temperature of the Comptonizing
medium increases, whereas the optical depth decreases. No CRSF is required by
the data, although the statistics at high energies does not allow us to
rule out the presence of a CRSF completely, especially if the CRSF is shifted to higher energies. 
The absorption corrected flux in 0.4-70\,keV
energy range is $\sim5\times10^{-10}\text{ergs\,cm}^{-2}\text{\,s}^{-1}$,
which corresponds to a luminosity of $\sim2.4\times10^{35}\text{ergs\,s}^{-1}$
for a distance of 2\,kpc.
\end{compactitem}
The observed pulsations, the luminosity and the hard spectrum of the
``off-states'' can only be explained if the emission is powered by the
accretion of plasma onto the magnetized neutron star. The absence of emission
lines in the ``off-state'' spectrum strongly suggests that the source is not
eclipsed, but instead exhibits an intrinsic drop in luminosity, hence in the accretion
rate. The timescale of the state transition makes it difficult, as argued by
\cite{kreyken_vela}, to explain the onset of the ``off-states'' with a sudden
decrease in wind density and/or velocity, and suggests a magnetospheric origin
of the state transition.

This agrees with the observed flux distribution. If the
``off-states'' were due to drops in the wind density, one would expect them to
contribute to the lower-flux \emph{tail} of the normal flux distribution. On
the contrary, they form a distinct low-flux \emph{peak} as observed in
Fig.~\ref{fig:hist}. The lognormal flux distribution is most likely caused, as
discussed by \cite{fuerst}, by the ``grinding'' of a clumpy wind by the
magnetosphere, while changes in the distribution parameters may be associated
with changes in the way the magnetosphere-plasma interaction proceeds.

As discussed by \cite{burnard83}, plasma generally enters the magnetosphere of
accreting pulsars according to various instabilities. These authors also
conclude that, for the observed luminosities and spin-periods typical of
bright accreting pulsars, the plasma mainly penetrates the magnetospheric
boundary via \rt instabilities. If the accretion rate decreases, the rotating
magnetosphere will inhibit accretion via \rt, therefore, for low-luminosity
pulsars with intermediate rotation rates, the \kh accretion channel dominates
\citep{burnard83}.
 
The ways through which the plasma can penetrate the magnetosphere have been reviewed more
recently by \cite{Bozzo:2008p2039}, who also provide estimates for
the leak rates of various mechanisms. For a system with parameters similar to
\vela in ``off-state'', the highest rate is expected to be provided by \kh
(see section 3.2.2 of \citeauthor{Bozzo:2008p2039} for the details). The accretion
luminosity is estimated in this case to be
\begin{eqnarray*}
& & L_{\rm KH} \simeq G M_{\rm NS} \dot{M}_{\rm KH}/R_{\rm NS}=\nonumber \\ 
& & 7.4\times10^{35}\eta_{\rm KH} R_{\rm M10}^3
(1+16 R_{\rm G10}/(5 R_{\rm M10}))^{3/2} 
\frac{\sqrt{\rho_{\rm i}/\rho_{\rm e}}}{
1+\rho_{\rm i}/\rho_{\rm e}} ~{\rm erg ~s}^{-1}\nonumber \\
\label{eq:lkh} 
\end{eqnarray*}
Here $R_{\rm G10}$ and $R_{\rm M10}$ are the capture and magnetosphere radius
respectively, in units of $10^{10}$\,cm; $\rho_{\rm i,e}$ are the densities
within and outside of the magnetosphere. According to \cite{Bozzo:2008p2039},
$\eta_{\rm KH}\sim 0.1$ and the density ratio is estimated to be between
\begin{equation*}
	\frac{\sqrt{\rho_{\rm i}/\rho_{\rm e}}}{
	1+\rho_{\rm i}/\rho_{\rm e}}=\begin{cases}
	\eta_{\rm KH} h^{-1} R_{\rm M10}^{3/2} P_{\rm s283.5}^{-1}\\
	0.1 \eta_{\rm KH} h^{-1} R_{\rm M10}^{1/2} v_{8}
\end{cases}
\end{equation*}
where $h$ is the fractional height of the area where the plasma and the
magnetic field coexist, in units of the total thickness of the \kh unstable
layer \citep{burnard83}, and $P_{\rm s283.5}$ is the spin-period in units of
283.5\,s. We assume a canonical neutron star radius of $R_{\rm
NS}\sim10\,{\rm km}$. In the case of \vela, for the observed ``off-state''
luminosity of $\sim2.4\times10^{35}\,{\rm erg\,s}^{-1}$, a magnetic field of
$B\ge2\times10^{13}$\,G is required, if the \kh unstable layer is relatively
thin ($h\sim0.05$), or $B\sim10^{14}$\,G, if $h\sim1$, as suggested by
\cite{burnard83}.

Evidence of such a high magnetic field in \vela, surprisingly stronger than
the one estimated from the CRSF energy, are extensively discussed in
\cite{dorosh_sub}. Here, we wish only to point out that the discrepancy still can
be cleared if the line formation region is located several kilometers
above the neutron star surface.
In fact, for the average observed luminosity of the normal state,
$\sim4\times10^{36}{\,\rm erg\,s}^{-1}$, one expects that an accretion column
with height up to $\sim10\,{\rm km}$ will arise
\citep{Lyubarsky1988,Doroshenko:2010p3661}. This implies a factor of ten decrease
in field strength at top of the column given that $B\sim B_0((R_{\rm
NS}+10)/R_{\rm NS})^{-3}\sim0.1B_0$. It is therefore sufficient to assume that
the accretion column exists and that the observed CRSF forms closer to the top
of the column to reconcile the strong magnetic field required by the observed
``off-states'' luminosity and the measured CRSF centroid energy.

The observed change in the high-energy pulse profile is explained under
the assumption that an accretion column does indeed exist. The sharp dip,
evident at high luminosity and softer energies, would then be due to the
eclipse of the polar cap by the accretion column. As the luminosity drops and
the column ceases to exist, the hard X-rays can pierce through and the polar
cap is observed directly: a pronounced peak is observed instead of a dip. In
other words, the high-amplitude peak, which appears in the ``off-state''
pulse profile at hard energies around pulse phase 0.75 can be attributed to
the direct emission from the polar cap. In this scenario, the accretion
stream would still absorb the soft X-rays so the dip is still observed at
lower energies. A similar scenario is discussed by \cite{kloch_exo} to
explain the pulse profile variations of EXO~2030+375 during outbursts.

\section{Conclusions} We presented an analysis of the ``off-states'' of \vela
observed during a 100\,ks \emph{Suzaku} observation of the source. For the
first time ever, we have been able to study and characterize the properties of
\vela during the so-called ``off-states'', detecting pulsations and measuring
the spectrum and flux distribution. We also confirm the presence of a CRSF at
25\,keV in the normal state of the source. We conclude that these
observational results strongly suggest that the emission is still powered by
accretion and that the drop in luminosity has a magnetospheric origin. The
observed X-ray luminosity of the ``off-state'' may be naturally explained in
the gated accretion scenario originally proposed by \cite{burnard83} and
recently investigated by \citep{Bozzo:2008p2039}, if the neutron star is
strongly magnetized and the plasma enters the magnetosphere via
Kelvin-Helmholz instabilities. To our knowledge this is the first time that
the theoretical considerations of magnetosphere-plasma interactions proposed
so far \citep{burnard83,Bozzo:2008p2039} find an observational confirmation.
Same interpretation might apply to several
other sources for which episodes similar to the ``off-states'' observed in \vela
have been reported \citep{paul05,kreyken10}.
\begin{acknowledgements}
	VD and VS thank the Deutsches Zentrums für Luft- und Raumfahrt (DLR) and Deutsche Forschungsgemeinschaft (DFG) for financial
	support (grants DLR~50~OR~0702 and SFB/Transregio~7: ``Gravitational Wave
	Astronomy''). VS was also supported the Russian Foundation for Basic Research
	(grant 09-02-97013-p-povolzh'e-a).
	\end{acknowledgements}
\bibliography{auto_clean} \end{document}